# Invalidity of prohibition of the perpetual motion engine of the second kind and the scenario of using these engines for prevention of the "thermal death" on the Earth: The transformation of power production and consumption into the circulation of heat

S.D. HAITUN

The leading researcher of the Institute for the History of Science and Technology
Russian Academy of Sciences
Moscow, Russian Federation <haitun@ihst.ru>

**The original Russian version**

CONTENTS

## Annotation

The paper is grounded on the author's book «"Thermal Death" on the Earth and the scenario how to prevent it. Part 1. Power engineering based on the heat circulation and perpetual motion engines of the second kind. Part 2. Perpetual motion engines of the second kind and invalidity of their prohibition» (Moscow, URSS, 2009, bibl. 571).

While consuming energy, we transform one of its forms into another and finally practically all the produced energy is dissipated as heat. When the consumption of energy became 0.1% (one estimation) or 1% (another estimation) of solar radiation reaching the Earth surface, the consequences of overheating would become catastrophic. With present-day rate of the growth of energy consumption, this would happen in 50–80 or 130–200 years. Deceleration of the energy consumption is also ruinous for mankind because it contradicts the evolution laws; or, speaking more precisely, deceleration is against the evolution vector directed to intensification of metabolisms.

As the way of preventing the “thermal death” the author suggests that we should transit to the thermocyclic power engineering based on using heat circulation and build “cold storage plants” which must collect the dissipated heat and transform it into forms of energy necessary to mankind. The increase of the energy consumption can be achieved due to acceleration of heat circulation.

The two types of “cold storage plants” are being conceived: firstly, thermal engines with coolers which efficiency does not exceed the Carnot efficiency, and secondly, thermal engines without coolers which efficiency is not limited by the Carnot efficiency, i.e. perpetual motion engines of the second kind. These engines are traditionally forbidden by the Second Law of thermodynamics. As a rule the dissipated heat has low temperature gradients. For this reason the efficiency of “cold storage plants” of the first type is too small and therefore they cannot be used as the basis of thermocyclic power engineering. This fact makes us consider the perpetual motion engines of the second kind. Having analyzed (excessively) numerous formulations of the Second Law of thermodynamics, the author has concluded that the prohibition of these engines is groundless.

From the author’s point of view this prohibition is a consequence of the following mistakes made by the founders of Thermodynamics. 1. Having in mind the concept of non-exterminable phlogiston, S. Carnot made a conclusion that the heat does not disappear but is only transferred from a warm body to a cold one and therefore a cooler is necessary for any thermal engine. However, today we know that while heat is consumed, it does disappear as heat and converts into other forms of energy. Hence, Carnot’s argumentation is wrong. 2. On the Earth there is a *tendency* to convert non-heat forms of energy into heat. This tendency is interpreted wrongly as a strict *law*. Conseqently, many scientists are sure that perpetual motion engines of the second kind are impossible and, moreover, they reject even – in spite of the facts – the possibility itself to use the dissipated heat as a source of “gratis” energy. 3. The Entropy Increase Law is wrongly reduced to the *Thermal* Entropy Increase “Law” while in Nature the *Total* Entropy Increase Law works. Due to groundlessness of the Thermal Entropy Increase “Law”, the thermal entropy can decrease and a compensation of converting thermal energy into mechanical work can be non-thermal. Therefore a cooler is not necessary for a thermal engine and perpetual motion engines of the second kind are possible.

## 1. The threat of mankind’s death due to the thermal pollution caused by the energy consumption as such

Being seriously concerned with global warming, the world community, as it seems to me, misses an important, and probably the most important, aspect of the problem. Nowadays the majority of experts believe that global warming is caused by industrial emissions of greenhouse gases into the atmosphere, mainly carbon dioxide $CO_2$ .Thereafter, the main efforts are directed to searching out means to decrease the concentration of $CO_2$ in the atmosphere.

This is, in my opinion, the mistake. Purification of the atmosphere would only delay the catastrophe for a short while. The point is that: consuming energy we only convert one form of energy into another. And eventually nearly all produced energy will dissipate as heat. As the result the biosphere undergoes overheating. If and when the mankind produces annually the amount of energy (which converts into heat) equal to Sun’s energy falling onto the Earth surface all developed forms of life will die. For sure, the biosphere will be unable to cope with the doubled amount of heat.

In 2003 the amount of produced energy was approximately 5000 times less than the amount of the Sun energy falling annually onto the Earth. However, the energy consumption doubles every 23.5–35 years (with the annual increase of 2–3%). It means that *with this rate of growth* the produced energy will become equal to the Sun energy in 285–430 years. Actually overheating becomes catastrophic when the produced energy reaches 0.1% (one estimation) or 1% (another estimation) of the Sun energy, i.e. in 50–80 or in 130–200 years correspondingly. This is the “thermal death”, the founders of thermodynamics talked about in the second part of the XIXth century. In the XXth century, this hypothesis as applied to the Universe was rejected. Nowadays the “thermal death” has become a real threat to the Earth.

It is essential that the thermal pollution emerges not because of the consumption of “bad” types of fuel but because of the energy consumption as such. Usage of “pure” types of fuel would not give us any advantage in this respect.

Until now the world community has not considered this problem. As for the global energy crisis, traditionally three its components are being perceived and we add the fourth one:

1) The exhaustion of hydrocarbon resources;
2) The injury of the security the countries deprived of the sufficient national power supply base;
3) The chemical and radioactive pollution by “unclean” energy sources;
4) The thermal pollution by the energy consumption as such, including usage of “pure” types of fuel.

As for the first two components, the world community reaction is sufficiently active. It seems that the response to the third component is inadequate, because in attempt to find substitutions for hydrocarbons the world community turns back to such “unclean” energy resources as coal and nuclear energy. And the world community, including ecologists, does not respond to the fourth component at all.

## 2. Deceleration of the energy consumption growth is against the evolution vector and therefore is ruinous

Few scientists have written about the threat of the thermal death caused by the exponential growth of the energy consumption. However, for some reason, they are sure that mankind is able to slow down the growth of energy production. Calls of that kind are accompanied by a powerful stream of similar calls motivated by the global ecological crisis. They say that the technological civilization and consumer society are doomed and have to turn back to the past. But I am sure that *a substantial deceleration of the energy consumption and of the consumption whatever is impossible at all*.

What brings us to this conclusion, is the actively developed during the last decades Universal Evolutionism **(**Big History). This branch of science considers non-organic, organic, and social evolution in universal vein. It is certain, that in the framework of universal evolutionism there exists the evolution vector with the several components [1. Sect. 4.8]:

1) Intensification of interchange of energy and matter;
2) Intensification and expansion of the energy and matter turnovers; and so on.

In the process of evolution the organic world did turn from lesser intensive metabolisms (fermentation) to more intensive ones (photosynthesis, respiration, and photorespiration) and consequently to macroergic compounds which play the central role in the work of cellular organelles responsible for providing cells with energy. Time after time the organic forms with more intensive energy metabolism became the winners in the evolution competition. For example, only animals with sufficiently large interchange of energy have skeletons. Mammals passed ahead of reptiles because their “red meat” saturated with oxygen (more exact with mitochondria) provides more intensive metabolism than the “white meat” of reptiles. And so on and so forth.

To obtain a clear understanding of the evolution vector, one should take into account not only a particular evolving system but its surroundings as well. For example, settled in caves live organisms regress in comparison with the terrestrial ones; but on the other hand metabolisms in the whole “cave” ecosystem become more intensive when compared to those of the times when the “cave” ecosystem was not populated by live forms.

The occurrence of a human being resulted in further intensification of metabolisms. When the whole history of mankind is considered as a macro history, when we abstract from the destinies of the single, inherently mortal societies, the history becomes the story on intensification of trade, economical, cultural, and other interactions of various elements and parts of the social world.

I believe that the evolution laws are compulsory laws of Nature like the law of Gravity, for example. We all know very well what happens if one jumps from a plane without a parachute. If mankind tries to hamper the energy consumption, i.e. to break the evolution law, it will die. In the past there were many societies which vanished from the face of the Earth because their behavior was “perpendicular” or not enough “parallel” to the evolution vector.

## 3. The author scenario: transition to thermocyclic power engineering based on the heat recirculation and perpetual motion engines of the second kind

I suggest that we not tempt our fate and (as a possible way of behavior) work out the measures that would allow us to prevent the “heat death” without slowing down rates of growth of the energy consumption. The idea sounds rather simply: is it possible to collect the dissipated heat so that to use it again and again?

The idea to collect the dissipated heat with its subsequent usage in power engines is not a chimera at all. This is precisely what geo- and geothermal power stations are doing. Thermal pumps, which are more and more widely used for building heating, collect heat from the Earth surface or atmosphere. Experimental oceanic power stations work by using the temperature difference between deep and surface layers of water (such a station has been installed on an old tanker).

Further, we will call the power stations which use the dissipated heat and thus cool the environment "cold factories", and the based on them power engineering we will call *thermocyclic*. If we manage to create a sufficient number of rather powerful and rather economical "cold factories", and if we manage to allocate them everywhere in the atmosphere, in the hydrosphere, and on the Earth crust, then, while collecting the dissipated heat, these "cold factories" will send almost all produced energy back to the process of energy circulation.

If we succeeded in realization of this project, the "cold factories" would transform the energy consumption into the heat-turnover; therefore not only the threat of the "heat death" would be removed but we would obtain an inexhaustible source of energy (the dissipated heat) and *the problem of exhaustion of energy resources would be obviated as well*. In this scenario the extraction of fossil fuels is reduced to minimum – only a small amount of energy (percents or even shares of percents), that is the lost in each cycle of the heat-turnover would have to be compensated by fossil fuels. Due to its acceleration, the heat-turnover would allow us to increase the energy consumption without destroying the ecological balance with the environment. It would happen in some way similar to how the organic world in the course of evolution has many times increased the annual consumption of energy and matter, while the biosphere mass has more or less been conserved.

This is in general the author's scenario of preventing the "heat death". The alternative scenario is based on deceleration of rates of growth of the energy consumption. *These two scenarios mean that within the hundred of years the mankind must change its style of life radically. At the same time the first scenario is directed along the evolution vector while the second one is against it*. Only few persons including the author, believe that the second scenario is ruinous. Neglecting the evolution vector, the overwhelming majority of the researchers unfortunately support the latter one.

But a scientific discussion differs from the real life. When it comes to the crucial aspects of human life, the mankind cannot rely on some point of view. One should not put all eggs in one basket. It is important to work out both scenarios so that in future to implement one of them or their combination.

Here we must say that, unfortunately, geo- and hydrothermal power stations have a small amount of energy resources. The efficiency of oceanic units is bounded above by the Carnot efficiency. Due to small temperature gradients in ocean, the upper limit of the efficiency is about 7%, but actually it does not exceed 2–3%. For the same reason the efficiency of thermal pumps is also low. Generally, the Second Law of thermodynamics (in its traditional interpretation) does not allow to increase the efficiency of the thermal engines with coolers.

According to the Second Law of thermodynamics in its traditional interpretation, energy units able to complete conversion of heat into mechanical work and thus of unlimited (not limited from above by Carnot efficiency) efficiency are impossible.

Consequently, all thermal power units must have a cooler which consumes part of heat what results in thermal entropy (the Clausius entropy) increase.

Hence, we have the treat of the “heat death” on one side, and the prohibition of perpetual motion engines of the second kind – on the other. In my view, the threat of the “heat death” is a powerful enough stimulus for the arguments in defense of perpetual motion engines of the second kind to be considered favorably. Since the prohibition of such engines is deduced from the Second Law of thermodynamics, we should look upon this Law more thoroughly.

## 4. Invalidity of prohibition of perpetual motion engines of the second kind: four mistakes made by the founders of thermodynamics

The situation is not as hopeless as it could be seen at the first sight. It must be impossible, from my point of view, that the evolution laws, which during a lot of time encouraged the organic world and mankind to intensify metabolisms and turnovers of matter and energy, stumbled suddenly over a law of physics (the Second Law of thermodynamics) which condemned the mankind to death by prohibition of the heat turnover. I am sure that the evolution laws and the laws of physics belong to the united and consistent set of laws of Nature. In this case the prohibition of the perpetual motion engine of the second kind “must” fail.

Based on this assumption, the considerable part of our book is devoted to the analysis of presented in scientific literature, (excessively) numerous formulations of the Second Law of thermodynamics and definitions of perpetual motion engines of the second kind. This analysis did lead me to a conclusion that the prohibition of such engines is invalid. In my opinion the prohibition is a consequence of several mistakes made by founders of Thermodynamics and Statistical Physics.

1. S. Carnot made a conclusion that any thermal engine has a cooler. This conclusion was based on the erroneous idea that the phlogiston does not ever disappear, i.e. the heat consumption (the phlogiston consumption) is similar to the energy consumption. Consuming energy we do not annihilate it, we only convert one form of energy into another. Carnot says that the phlogiston consumption means that the phlogiston passes from a warmer body to a cooler one. According to Carnot, such a cooling element must be in any thermal engine.

The classics of Thermodynamics who came after Carnot, abandoned the phlogiston theory but they admitted that a thermal engine must have a cooler. In this way they supported the prohibition of perpetual motion engines of the second kind, i.e. thermal engines without cooling elements. This conclusion is to the highest degree strange because *heat ceases to exist as a heat* being converted into other forms of energy. In Carnot’s terms we may state that while consuming phlogiston we annihilate it, whereupon Carnot’s argumentation is invalid.

2. R. Clausius, W. Thompson and other followers of S. Carnot, having abandoned the phlogiston theory, did not correct his mistake because they worked only with such thermal engines which required coolers. Those thermal engines required coolers because they had the following features:

1) they were cyclic;
2) they used a homogeneous and one-phase working substance.

Indeed, when such a working substance returns to its initial state a part of heat received from a heater has to be given to a cooler. Meanwhile non-cyclic thermal engines and cyclic thermal engines with a two-phase working substance do not require coolers.

An example of a non-cyclic thermal engine without a cooler is a jet engine working in vacuum. In this case the combustion residues are not cooled because the free gas expansion has an isothermal character. This is true for an ideal gas because its internal energy depends only on temperature [3. C. 148] and, with a sufficient accuracy, this is also valid for real gases [4. C. 44].

During the last three decades several authors [5–12] have independently showed that in cyclic thermal engines with a two-phase working substance (gas–liquid) the process of its returning into the initial state can be realized without passing some heat to a cooler but by returning this heat to a heater. Such engines do not need coolers and their efficiency is not limited by the Carnot efficiency.

3. The Entropy Increase Law is reduced incorrectly to the *Thermal* Entropy Increase "Law" while in Nature the *Total* Entropy Increase Law acts. If one accepts that the thermal entropy increase "law" is not valid than one should admit that the thermal entropy can decrease, and a thermal engine can convert the heat into mechanical work without a cooler, and the corresponding compensation can be non-thermal.

4. Usually non-thermal forms of energy are converted into heat. This *tendency* is interpreted incorrectly as a strict *law*. And therefore some scientists are sure that perpetual motion engines of the second kind are impossible, and contrary to reality they reject even the possibility of using a dissipated heat as a gratis source of energy itself.

## 5. Clarification of the Second Law of thermodynamics

The key point of our discussion is the Second Law of thermodynamics. Numerous existing in scientific literature formulations of the Law say that its content is rather blurred. In our book we give 48 various formulations, but actually there are many others. This diversity contrasts with the formulations of the law of energy conservation which are practically identical in all scientific sources. A great number of formulations of the Second Law of thermodynamics say that there is no clear understanding of its meaning.

The main goal of our book is to shell a kernel of the law of Nature which is hidden in all these formulations. In doing this we will be keeping in mind that we want to find out whether perpetual motion engines of the second kind are actually forbidden or not.

It has been written many times that various formulations of the Second Law of thermodynamics are identical. Certainly, all numerous formulations cannot be identical and at the same time they cannot be different. These formulations form a complicated set. However, this set can be ordered to some extent.

The made in the book analysis has shown that there are five assertions **(**points**)** in what is called the Second Law of thermodynamics. One of them is not valid. The point are:

1) in thermal processes the temperatures are equalized, the heat passes from a warmer body to a cooler one;
2) there is an asymmetry between processes of converting non-thermal forms of energy into the heat and opposite processes of converting the heat into other forms of energy. The processes of the first type do not require any compensation, while the others require it;
3) in the case of an equilibrium (reversible) process we have the definition of the thermal entropy $dS = dQ/T$;
4) there acts the total entropy increase law;
5) there acts the thermal entropy increase "law".

The first assertion has nothing to do with perpetual motion engines of the second kind because in such engines processes of converting the heat into other forms of energy are not only thermal. The situation with third assertion is the same because the processes occurring in perpetual motion engines of the second kind are irreversible. Therefore only the second, forth and fifth assertions can be relevant to such engines.

First, let us make more precise what is "seditious**"** with the perpetual motion engines of the second kind. The problem is in the following: in these engines the conversion of heat into other forms of energy *is not compensated by thermal means* and, as a result, the thermal entropy decreases.

Let us come back to the Second Law of thermodynamics. There are two types of formulations of the second assertion. The formulations of the first type say that converting the heat into other forms of energy does require a *thermal* compensation. Consequently they reflect the "law" of increase of the *thermal* entropy. The formulations of the second type say about a compensation assuming (implicitly) that it can be *non-thermal*. And in this way they reflect the law of increase of the *total* entropy.

In other words, according to the thermal entropy increase "law", the processes of converting the heat into other forms of energy require a sufficient thermal compensation. In this case the perpetual motion engines of the second kind are forbidden. But according to the total entropy increase law such processes do not require a thermal compensation. It is sufficient that the total entropy increases. Consequently, the perpetual motion engines of the second kind are possible.

In out book we present arguments in defense of the total entropy increase law and make a conclusion that the thermal entropy increase "law" does not work. This conclusion is supported by nine examples of converting the heat into other forms of energy without a thermal compensation, i.e. processes occurring with the thermal entropy decrease.

## 6. Converging flow of ideal gas: compensation of a decrease of the thermal entropy by an increase of the non-thermal entropy

We present here only one example – a converging horizontal flow of ideal gas with constant specific heat (say, in a conical tube). Due to "geometrical" reasons such a flow accelerates along the streamline. For simplicity we assume the gas flow to be horizontal in order to exclude the influence of the Earth gravitational field. Such a flow is described by the Bernoulli equation [13. C. 37]

$$\frac{v^2}{2} + c_p T = const, \qquad (1)$$

where $v$ is the flow velocity, $c_P$ is the specific heat at constant pressure, $T$ is the temperature. Equation (1) can be rewritten in the form

$$\frac{\gamma}{\gamma - 1}\frac{p}{\rho} + \frac{v^2}{2} = const, \qquad (2)$$

where $\gamma = c_P / c_V$, $c_V$ is the specific heat at constant volume, $p$ is the pressure, $\rho$ is the gas density. Equation (2) shows that in an accelerated gas flow not only the temperature decreases but the pressure as well (as it is known, this effect provides the lift of the wing).

According to Eq. (1), acceleration of the gas flow is accompanied by its cooling. The entropy of ideal gas at constant specific heat is

$$S = -N \ln p + N c_P \ln T + const, \qquad (3)$$

where $N$ is the number of particles [3. C. 151]. One can see from (3) that when the gas flow is cooling its thermal entropy is decreasing, but it is compensated by an entropy increase caused by a decrease of flow pressure, i.e. by a non-thermal way, which was to be proved.

## 7. The thermal entropy vs. the total entropy

Having contradistinguished the two notions – the total and the thermal entropy – we discuss the total entropy increase law which works and the thermal entropy increase "law" which does not work. It seems that everything is clear: if we distinguish the total and the thermal energies we should distinguish the total and the thermal entropy as well.

But we do not see such a contradistinction in the scientific literature. The notion of entropy is still fuzzy. This fuzziness is manifested in various ways. Say, the probability $W$ in the Boltzmann principle $S = k \ln W$ is still called the *thermo*dynamic probability though it is clear that in general case the probability of a state of a macroscopic physical system (the number of microstates by which the given macroscopic state can be realized) is defined not only by thermal phenomena. In general case a physical system for which the probability of a state is characterized by $W$ is not a thermodynamic system.

Confusion of notions of the total and thermal entropy is typical for all modern courses of physics (70), which I studied while working on the book. The terms and system thought are mainly oriented on the *thermal* entropy. While talking with physicists one can hear that there is no non-thermal entropy at all. All this is far from being harmless because it is the confusion of the total and thermal entropy where the ban on perpetual motion engines of the second kind is rooted.

## 8. The definition of the total and thermal entropy in the general irreversible case

Strange, as it may seem, but until now there is no quantitative definition *of the*

*total entropy in the general irreversible case*. (Probably for just this reason scientists do not distinguish the total and thermal entropies.) I think that the total entropy can be defined as entropy of the distribution of the *total energy* of a system. More precisely, in general case we may introduce the total energy density distribution over the three-dimensional volume (the space where we live) and over the three-dimensional space of the speed of propagation of energy, or in other words, the flux density of the total energy $U(r,v)$ :

$$\int U(r,v)drdv = U \,, \tag{4}$$

$$\int U(r,v)dv = U(r) \,, \tag{5}$$

where $U$ is the total energy of a given material system, $U(r)$ is the spatial density of the total energy, $r$ is a three-dimensional radius-vector.

Let us pass from $U(r,v)$ on to $u(r,v)$ so that the distribution (5) renormalize to unity:

$$\int u(r,v)drdv = 1 \,. \tag{6}$$

Such a normalization allows us to interpret $u(r,v)$ as the *probability density* of the total energy distribution in the six-dimensional space of coordinates and speeds, mentioned above.

We define the total entropy of a material system as

$$S = -k\int u(r,\, v)\ln u(r,\, v)drdv \,, \tag{7}$$

where $k$ is the Boltzmann constant.

The definition (7) generalizes the definition of the Boltzmann entropy

$$S = -k\int f^1(r_1,\, p_1)\ln f^1(r_1,\, p_1)dr_1dp_1 \,. \tag{8}$$

Here $f^1(r_1,\, p_1)$ is the distribution function of a system in the six-dimensional space of coordinates $r_1$ and momentums $p_1$ of one particle. The distribution function means a number of particles per unit volume of this six-dimensional phase space.

The normalization (6) of the distribution function $u(r,v)$ means that the area under the curve $u(r,v)$ is equal to unity. As in statistical physics, let us introduce the average value $u(r,v)_{MEAN}$ and the distribution width $(\Delta r\Delta v)_U$ as correspondingly the height and the width of a rectangle of unit area**:**

$$u(r,v)_{MEAN}(\Delta r\Delta v)_U = 1 \,. \tag{9}$$

The quantity $(\Delta r\Delta v)_U$ may be called the *phase volume* of the distribution $u(r,v)$. It is easy to show that the entropy (7) takes the form

$$S = k\ln(\Delta r\Delta v)_U \,. \tag{10}$$

Indeed, taking into account

$$\ln[u(r,v)_{MEAN}] \approx [\ln u(r,v)]_{MEAN} \,, \tag{11}$$

i.e.

$$\ln\int u(r,v)u(r,v)drdv \approx \int u(r,v)\ln u(r,v)drdv \,, \tag{12}$$

we obtain

$$S = k\ln(\Delta r\Delta v)_U = k\ln\frac{1}{u(r,v)_{MEAN}} =$$

$$= -k\ln[u(r,v)_{MEAN}] \approx -k[\log u(r,v)]_{MEAN} = -k\int u(r,v)\ln u(r,v)drdv,$$

(13)

i.e. the definition of the total entropy (7).

The transformations similar to those which have led us from (10) to (13) are used usually in statistical physics [3. C. 40–41].

Comparing (10) with the entropy $S = k \ln W$, we conclude that the quantity $(\Delta r \Delta v)_U$ can be interpreted as the macro-probability of an energy state of a system under consideration:

$$S = -k \int u(r,\, v) \ln u(r,\, v) dr dv = k \ln (\Delta r \Delta v)_U = k \ln W_U \, .$$

(14)

We should note that the entropy (7) has some advantages as compared to the Gibbs entropy

$$S = -k \int \rho(q,\, p) \ln \rho(q,\, p) dq dp \, , \qquad (15)$$

where $\rho(q,\, p)$ is the distribution function of a system in $\Gamma$-space. Gibbs' entropy is a *mechanical* quantity, so it can be used only for systems which are described by coordinates and momentums; and it cannot be used for non-mechanical systems.

R.Clausius defined the thermal entropy by

$$dS = \frac{dQ_{EQUIL}}{T} \, , \qquad (16)$$

$$\Delta S = S_2 - S_1 = \int_1^2 \frac{dQ_{EQUIL}}{T} \, . \qquad (17)$$

This definition is valid only in (quasi)equilibrium cases. Unfortunately, there is no definition of the thermal entropy for non-equilibrium processes when we need it. The definition of the total entropy (7) allows us to fill up the gap. For this end it is just sufficient to replace the probability density of the total energy distribution $u(r, v)$ by the probability density of the thermal energy distribution $q(r, v)$:

$$S_Q = -k \int q(r,\, v) \ln q(r,\, v) dr dv \, ,$$

(18)

$$\int Q(r, v) dr dv = Q \, , \qquad (19)$$

$$\int Q(r, v) dv = Q(r) \, ,$$

(20)

$$\int q(r, v) dr dv = 1 \, ,$$

(21)

where $Q$ is the amount of the thermal energy in a given material system, $Q(r)$ is the spatial density of the thermal energy.

Similar to how it was done for the total entropy, we introduce the "phase volume" $(\Delta r \Delta v)_Q$ for the distribution $q(r, v)$. Then the thermal entropy (18) takes the form

$$S_Q = k \ln (\Delta r \Delta v)_Q \, . \qquad (22)$$

Comparing (23) with Boltzmann's principle $S = k \ln W$, we see that the quantity

$$(\Delta r \Delta v)_Q = W_Q \qquad (23)$$

can be interpreted as the macro-probability of a (thermal) state of a given system and the quantity

$$\frac{dE}{kd(\ln W_Q)} = \frac{dE}{dS} = T \qquad (24)$$

is interpreted as its temperature.

Since the Clausius thermal entropy is defined only for equilibrium processes, we compare the thermal entropy (18) (defined in general case of equilibrium or non-equilibrium system) with it for equilibrium processes. R.Reif [14. C. 153] studied an equilibrium system which absorbed a small amount of heat $Q$. He calculated a change of Boltzmann's entropy $S = k \ln W$. Here we present his calculations replacing $W$ by $W_Q$.

With a great probability the initial and final energies of an equilibrium system $A$, which absorbed a small amount of heat $Q$, are equal to averaged values $\bar{E}$ and $\bar{E} + Q$ correspondingly. Using the Taylor series we obtain

$$\ln W_Q(\bar{E} + Q) - \ln W_Q(\bar{E}) =$$

$$= \left( \frac{\partial \ln W_Q}{\partial E} \right) W_Q + \frac{1}{2} \left( \frac{\partial^2 \ln W_Q}{\partial E^2} \right) W_Q{}^2 + \ldots = \beta W_Q + \frac{1}{2} \frac{\partial \beta}{\partial E} W_Q{}^2 + \ldots \tag{25}$$

for the change of the number of possible states $W_Q(E)$ of the system $A$ caused by the absorbed heat $Q$.

Since $Q$ is small, the thermodynamic temperature $T = (k\beta)^{-1}$ of the system $A$ remains almost unchanged. Therefore, we can neglect the term proportional to $\partial\beta / \partial E$. As a result we have the change of $\ln W_Q(E)$

$$\Delta(\ln W_Q) = \frac{\partial(\ln W_Q)}{\partial E} W_Q = \beta Q . \tag{26}$$

Thus, we see that if an equilibrium system absorbs a small amount of heat $Q$ its thermal entropy $S = k \ln W_Q$ changes by a small quantity $\Delta S = Q / T$. This proves that in an equilibrium case the thermal entropy (18) coincides with the Clausius thermal entropy.

So the author definitions of the total (7) and thermal (18) entropies generalize the existing definitions of entropy in the general irreversible case. We give these definitions following general rules of physics: physical entropy is considered as entropy $-k \int f(x) \ln f(x) dx$ of some distribution $f(x)$, describing physical system. The Boltzmann entropy (8) and the Gibbs entropy (15) have indeed such structure.

Having a greater generality than any other statistical entropy defined before, the total entropy (7) can be used for any material system (reversible or irreversible). In our investigation the main advantage of the total entropy (7) and the thermal entropy (18) is that they allow us to distinguish them clearly. Consequently, we can distinguish the total entropy increase law, which actually works, from the thermal entropy increase "law", which is invalid. As a result, prohibition on perpetual motion engines of the second kind is failed and the way to thermocyclic power engineering becomes open.

## 9. Prospects of the transition to the thermocyclic power engineering based on perpetual motion engines of the second kind

The fact that the physical laws do not prohibit perpetual motion engines of the second kind, does not mean that the creation of a sufficiently powerful, economical, and environmentally safe engines of this type, is actually possible. For example, thermonuclear fusion is allowed by the physical laws; however more than half a century work on creation of the thermonuclear power unit has been carried out with no result.

Yet we do not know how perpetual motion engines of the second kind would be created. In the last chapter of our book six projects of such engines are considered. The author has selected these projects as the most reliable from dozens published in literature and Internet. It does not mean that other projects of perpetual motion engines of the second kind are not worth a damn. Perhaps they are reasonable but the author does not undertake to judge them.

Unfortunately, due to the ban on perpetual motion engines of the second kind put by Big Science, this area of research is pushed to periphery of science. Such a situation made scientists working on projects of perpetual motion engines of the second kind stew in their own juice, and as a result the quality of their publications is as a rule extremely low. For this reason consideration of such projects is rather difficult. It is difficult to sort the wheat from the chaff.

Although the author is positively disposed to the idea of perpetual motion engines of the second kind, his attitude to each such project (with rare exceptions) is quite skeptical because information on developments sounds very often apocryphal. Here I am going to talk about one of such projects.

Let us again consider a converging flow of ideal gas (see Section 6). Acceleration of this flow is accompanied by its cooling. Put a converging tube upwind. The air in the tube will be accelerated due to the dissipated heat. We can make a cold storage plant by installing a turbine in the end of the tube. A group of Russian inventors, with I.S. Orlov in charge [15–17], offered to put the tube and turbine in a single enclosure. Their installation looks at the drawings as a pot-bellied bomb hanging along the air flow and taking it inward annular aperture. The inventors put the cascade of three Laval nozzles inside this unit. The turbine is put in the last nozzle. The projection “of the front view” (across the wind) of each nozzle is a ring, and the projection “of the side view” (along the wind) lets a nozzle remain the Laval nozzle.

If the inventors have calculated correctly, the air flow in such a unit will be accelerated almost up to the speed of sound. Their installation is protected by patents, but I have not seen it embodied in metal.

If a reader has a necessary equipment (the author does not have it), he can make experimentum crucis. For example, a converging tube can be made of plasitc film fixed on a wire cage.

Perhaps, a wind is not necessary for Orlov’s installation; I suppose that after starting with a “starter”, the unit itself will suck the air because acceleration of the converging air flow is accompanied by not only the temperature decrease but the pressure decrease as well.

I think that aggregates of that kind could be adjusted for water where, in my opinion, they would have larger efficiency.

I have described Orlov’s installation as an example. My goal is not in presenting projects of perpetual motion engines of the second kind which are ready for production but in reversing the negative attitude of Big Science to the very idea of such engines. I hope that, if and when large Energy Enterprises together with scientists undertake the task of selecting the most prospective projects of perpetual motion engines of the second kind the results, as I expect**,** would not make a long wait.

## APPENDIX. Contents of the book «"The heat death" on the Earth and the scenario how to prevent it. Part 1. Power engineering based on the heat circulation and perpetual motion engines of the second kind. Part 2. Perpetual motion engines of the second kind and invalidity of their prohibition» by S.D. Haitun (Moscow, URSS, 2009, bibl. 571)